\documentclass[a4paper,11pt]{article}
\pdfoutput=1
\usepackage{jheppub}

\title{An ansatz for one dimensional steady state configurations}

\author[a]{H. -C. Chang,}
\author[a]{A. Karch}
\author[b]{and A. Yarom}

\affiliation[a]{Department of Physics, University of Washington, Seattle, Wa, 98195-1560, USA}
\affiliation[b]{ Department of Physics, Technion, Haifa 32000, Israel }
\emailAdd{hanchih@uw.edu}
\emailAdd{akarch@uw.edu}
\emailAdd{ayarom@physics.technion.ac.il}

\abstract{We conjecture a universal formula for the heat current of a steady state connecting two asymptotic equilibrium systems in $d$-dimensional conformal field theories. Our proposal is verified by comparing it to exact expressions in 1+1 dimensions and linear hydrodynamics as well as numerical simulations in an Israel-Stewart like theory of second order viscous hydrodynamics.}

\begin{document}

\maketitle

\section{Introduction}

While statistical mechanics provides us with powerful tools to analyze systems in equilibrium, most of the world around us is far from this idealized situation. Near thermal equilibrium, one can use hydrodynamics to analyze the dynamics of small, long wavelength fluctuations. But to understand far from equilibrium physics is a notoriously challenging problem. An interesting yet potentially tractable class of non-equilibrium configurations is represented by steady state flows. Examples of steady state flows are an electric current in a conductor driven by an external electric field or a heat current driven by a temperature gradient. These configurations can be described by a time-independent configuration, but neither of these corresponds to equilibrium. In both cases entropy is produced constantly and needs to be absorbed by the battery or heat bath in order to maintain the steady flow.

In their recent papers \cite{bd1,bd2} Bernard and Doyon showed that in the highly constrained class of $1+1$ dimensional relativistic conformal field theories (CFTs) one can derive a universal formula for the heat flow in a steady state. The configuration they consider corresponds to an initial condition which interpolates sharply between two thermal states very far apart (somewhat reminiscent of a spatial quench). Indeed, if two initially equilibrated systems with given temperature and chemical potential are brought into thermal contact at time $t=0$, a steady state forms in the central region. Such a steady state will form as long as the system is infinite in extent and it will posses a non-trivial heat flow that is uniquely determined by the temperatures and chemical potentials of the asymptotic heat baths as well as the central charge and the current algebra level of the conformal field theory. Unlike the examples of steady states described in the previous paragraph, the steady state of Bernard and Doyon is actually dissipation free and the heat transport is dominated by flow rather than diffusion. Given their beautiful result, one may wonder whether a similarly universal answer can be obtained in conformal field theories in higher dimensions. This is the question we wish to address in this paper.

We start our analysis by re-deriving the Bernard-Doyon (BD) result in $1+1$ dimensions, using only energy conservation and the conformal form of the stress tensor. Our analysis allows us to extend the BD result in several interesting directions. We can see that the appearance of the steady state as well as its properties are insensitive to the details of the initial state. As long as the system is held at temperature $T_L$ and chemical potential $\mu_L$ at negative spatial infinity, $x \rightarrow - \infty$, and $T_R$ and $\mu_R$ at spatial infinity, $x \rightarrow + \infty$, then the universal steady state will develop at late times. This result is not just a property of the particular initial condition considered in \cite{bd1,bd2}. Thus, it should be much easier to test the steady state prediction experimentally. We can also see in detail how the steady state develops. Two wavefronts, moving at the speed of light, push outward into the asymptotic equilibrium region, leaving behind a region dominated by the steady state in the center. Moreover, it is straightforward to generalize the BD result to an arrangement where heat and charge currents are injected into the system at spatial infinity. 
Last but not least, we extend the analysis to setups where the underlying conformal field theory is intrinsically chiral, that is, we allow for separate left and right moving central charges $c_+$ and $c_-$ as well as separate current algebra levels $k_+$ and $k_-$.

Given the experience gained from this re-analysis of a two dimensional system we conjecture a simple ansatz for the behavior of the final steady state heat flow in higher dimensional conformal theories set up so that the dynamics depend on only one spatial coordinate. Our ansatz relies on the assumption that at late times the system settles to a configuration where the relevant components of the stress tensor are comprised of two wavefronts (which we refer to as ``steps'') that push into the asymptotic equilibrium heat bath, connected by a smooth velocity and pressure profile. Using the ``steps ansatz'' we can derive universal expressions for the heat current in the resulting steady state realized in the central region between the two steps. Such a steady state will form if the heat baths are infinitely far apart, or if the system is studied at times which are large compared to the mean free path of the system (in units of the speed of sound) but small compared to the system size. In order to show that our ansatz can actually be reached in realistic systems, we analyze numerically the evolution of various initial conditions within the framework of viscous relativistic hydrodynamics. Relativistic hydrodynamics will be a good description of the dynamics of the system as long as the width of the initial temperature profile is larger than the microscopic scales in the system (which in a CFT are simply set by temperature and chemical potential). We find that the hydrodynamic description indeed settles to our step ansatz. As the universal answer derived from the step is insensitive to the details of the hydrodynamic description (e.g. the shear viscosity), we believe it is valid much more generally. Unfortunately we are currently not able to prove it rigorously. Our proposal however should be straightforward to test experimentally.

The paper is organized as follows. In section \ref{S:2d} we provide a rederivation of the two dimensional result together with generalizations to non-trivial initial states, non-vanishing injected currents and chiral systems. In section \ref{S:higherd} we propose a generalization to higher dimensions. We first show that at the level of linearized hydrodynamics, the two dimensional analysis carries through in a straightforward manner. We then give an ansatz, inspired by the two dimensional analysis for a stress-energy profile in terms of two steps pushing into the asymptotic equilibrium regions. This ansatz allows us to derive universal expressions for the steady state in terms of the properties of the asymptotic heat baths. We then discuss how our ansatz may be generalized to non relativistic systems and conclude by numerically solving the non-linear equations of relativistic viscous fluid dynamics and confirm that, within the validity of hydrodynamics, initial configurations indeed settle, at late times, to our two step ansatz. We discuss our findings in section \ref{sec:conclude}.

\section{The two dimensional steady state}
\label{S:2d}

Consider a two dimensional CFT (conformal field theory) with left and right central charges $c_+$ and $c_-$, and a holomorphic and antiholomorphic current with levels $k_+$ and $k_-$ respectively.\footnote{Note that in the literature it is common to use $L$ and $R$ subscripts instead of $+$ and $-$. Here $L$ and $R$ subscripts are used exclusively for the left and right heat bath.}
Writing the flat space line element in the form
\begin{equation}
	ds^2 = dz d\bar{z}\,,
\end{equation}
conformal invariance, which requires that the stress tensor $T_{\mu \nu}$ is traceless, implies that
\begin{equation}
T_{z \bar{z}} =0.
\end{equation}
Conservation of $T_{\mu \nu}$ then requires that $T_{zz}$ is only a function of $z$, whereas $T_{\bar{z} \bar{z}}$ is only a function of $\bar{z}$. Hence we can write
\begin{equation}
	T_{\mu\nu}=\begin{pmatrix} T_+(z) & 0 \\ 0 & T_-(\bar{z}) \end{pmatrix}\,.
\end{equation}
Similarly,
\begin{equation}
	J_{+\,\mu} = \begin{pmatrix} j_+(z), & 0 \end{pmatrix}
	\qquad
	{J}_{-\,\mu} = \begin{pmatrix} 0, & {j}_-(\bar{z}) \end{pmatrix} 	\,.
\end{equation}
In Cartesian coordinates, $ds^2 = -dt^2 + dx^2$, we have
\begin{equation}
	T^{\mu\nu} = \begin{pmatrix}
		T_+(t+x) +{T}_-(-t+x), &   {T}_-(-t+x) - T_+(t+x) \\
		T_-(-t+x) - {T}_+(t+x),  & T_+(t+x) + {T}_-(-t+x)
	\end{pmatrix}
\label{cartesianstress}
\end{equation}
and
\begin{equation}
	J_{+}^{\mu} = \begin{pmatrix} -j_+(t+x), & j_+(t+x) \end{pmatrix}
	\qquad
	J_{-}^{\mu} = \begin{pmatrix} {j}_-(-t+x), & {j}_-(-t+x) \end{pmatrix}\,.
\end{equation}

Now suppose we have an initial value problem were we specify the value of the stress tensor and currents on some initial time slice $t=0$ and time independent boundary conditions at spatial infinity $x=\pm\infty$. Due to the holomorphic structure of the stress tensor and current, the boundary conditions completely specify the system at future infinity. Suppose that
\begin{align}
\begin{split}
\label{E:BCs}
	T^{\mu\nu}(t,x)\Big|_{x\to\pm\infty} &= \begin{pmatrix} \epsilon_{R/L} & Q_{R/L} \\ Q_{R/L} & \epsilon_{R/L} \end{pmatrix} \\
	J_{-}^{\mu}\Big|_{x\to\pm\infty} &= \begin{pmatrix} \rho^-_{R/L}, & \rho^-_{R/L} \end{pmatrix} \\
	J_{+}^{\mu}\Big|_{x\to\pm\infty} &= \begin{pmatrix} \rho^+_{R/L}, & -\rho^+_{R/L} \end{pmatrix}\,,
\end{split}
\end{align}
where the subscript $R$ corresponds to $x\to\infty$ and $L$ corresponds to $x\to-\infty$.
Then
\begin{subequations}
\label{E:main}
\begin{equation}
	T^{\mu\nu}\Big|_{t\to\infty} = \begin{pmatrix}
		\frac{1}{2} ((\epsilon_L+\epsilon_R) + (Q_L-Q_R)) & \frac{1}{2} ((\epsilon_L-\epsilon_R)+(Q_L+Q_R))  \\
		\frac{1}{2} ((\epsilon_L-\epsilon_R)+(Q_L+Q_R))  & \frac{1}{2} ((\epsilon_L+\epsilon_R) + (Q_L-Q_R))\,.
	\end{pmatrix}
\end{equation}
and
\begin{equation}
\label{steadystate}
	J_{-}^{\mu}\Big|_{t\to\infty} = \begin{pmatrix}
		\rho^-_{L}, & \rho^-_{L}
	\end{pmatrix}
	\qquad
	J_{+}^{\mu}\Big|_{t\to\infty} = \begin{pmatrix}
		\rho^+_{R},  & -\rho^+_{R}
	\end{pmatrix}
\end{equation}
\end{subequations}
That is, at any given (fixed) position $x$, at times larger than $|x|/c$, where $c$ is the speed of light, the system will be in a steady state with the currents completely specified by  the asymptotic energy and charge densities, eq.~(\ref{E:main}). As advertised in the introduction, the system is described by two wavefronts (steps) in temperature pushing out into the asymptotic equilibrium regions at the speed of light, leaving behind a stationary state with universal behavior.

To rewrite the properties of the steady-state configuration in terms of the temperature and chemical potential of the asymptotic heat baths, we identify our expression for the currents at the spatial boundary \eqref{E:BCs} with that of a system in thermal equilibrium. The thermodynamic description of a generic two dimensional conformal field theory has been succinctly summarized in \cite{Loganayagam:2012zg}. We find
\begin{subequations}
\label{E:Thermal}
\begin{equation}
	T^{\mu\nu}_{eq} = \epsilon (2 u^{\mu} u^{\nu} + \eta^{\mu\nu}) + \tilde{\epsilon} (u^{\mu} \tilde{u}^{\nu} + u^{\nu} \tilde{u}^{\mu})
\end{equation}
and
\begin{equation}
	J^{\mu}_{+\,eq} = \rho^+ \left( u^{\mu} - \tilde{u}^{\mu} \right)
	\qquad
	J^{\mu}_{-\,eq} = \rho^- \left( u^{\mu} + \tilde{u}^{\mu} \right)
\end{equation}
where
\begin{align}
\begin{split}
	\epsilon &= \frac{\pi}{12}(c_-+c_+)T^2 + \frac{1}{2\pi}\left(k_-(\mu^-)^2 + k_+(\mu^+)^2\right) \\
	\tilde{\epsilon} &= \frac{\pi}{12}(c_- - c_+)T^2 + \frac{1}{2\pi}\left(k_-(\mu^-)^2 - k_+(\mu^+)^2\right) \\
	\rho^+ & = \frac{k_+ \mu^+}{\pi} \qquad
	{\rho}^- = \frac{k_- \mu^-}{\pi} \qquad
	\tilde{u}^{\mu} = \epsilon^{\mu\nu}u_{\nu}\,.
\end{split}
\end{align}
\end{subequations}
Before proceeding it is worthwhile to point out that if $c_+ \neq c_-$ then the system naturally has a heat current even in thermal equilibrium. Also, even if $k_+ \mu^+ = k_-\mu^- = \frac{k \mu}{4}$ then we find that the ``vector'' current satisfies $J_+^{\mu} + J_-^{\mu} = \begin{pmatrix} \frac{k\mu}{2\pi}, & 0 \end{pmatrix}$ but the ``axial'' current transports charge $J_+^{\mu} - J_-^{\mu} = \begin{pmatrix} 0,  & -\frac{k\mu}{2\pi} \end{pmatrix}$

We choose our lab frame so that at the the left bath (at $x\to-\infty$) we have $u^{\mu} = (1,0)$, $T=T_L$ and $\mu^{\pm} = \mu^{\pm}_L$ and at the right bath we have $u^{\mu} = \gamma(1,\beta_R)$, $T=T_R$ and $\mu^{\pm} =\mu^{\pm}_R$. Rewriting the asymptotic expansion \eqref{E:BCs} in terms of the physical parameters associated with the left and right heat bath \eqref{E:Thermal}, we find
\begin{align}
\begin{split}
\label{E:TandJinSS}
	T^{tt}\Big|_{t\to\infty} &=
		\frac{\pi}{12}\left(c_- T_L^2+ c_+ T_R^2 \frac{1-\beta_R}{1+\beta_R}\right) +\frac{1}{2\pi}\left(k_- \mu^-_L + k_+ \mu^+_R \frac{1-\beta_R}{1+\beta_R} \right) \\
	T^{tx}\Big|_{t\to\infty} &=
		\frac{\pi}{12}\left(c_- T_L^2 -  c_+ T_R^2 \frac{1-\beta_R}{1+\beta_R} \right)+\frac{1}{2\pi}\left(k_- \mu^-_L - k_+ \mu^+_R \frac{1-\beta_R}{1+\beta_R} \right) \\
	J_-^t\Big|_{t\to\infty} = J_-^x\Big|_{t\to\infty} & = \frac{k_- \mu^-_L}{\pi} \\
	J_+^t\Big|_{t\to\infty} = -J_+^x\Big|_{t\to\infty} &= \frac{k_+ \mu^+_R}{\pi} \sqrt{\frac{1-\beta_R}{1+\beta_R}}
\end{split}
\end{align}
This is our final result for the steady state currents in the most general case of a chiral CFT with non-vanishing heat and electric currents injected at infinity.

Taking various limits of \eqref{E:TandJinSS} allows us to gain some insight into the behavior of the steady-state configuration and compare \eqref{E:TandJinSS} with the results in the literature,
\begin{itemize}
\item{{\bf Zero charge}}
	If we take $\mu^+_R=\mu^-_L=0$ and $\beta_R=0$ we find
	\begin{equation}
		T^{tx} = \frac{\pi}{12} \left(c_- T_L^2 - c_+ T_R^2\right)
		\qquad
		J_{\pm}^{\mu}=0
	\end{equation}
in perfect agreement with \cite{bd1,bd2}.
\item{{\bf Chemical potential driven}}
	If we take $\beta=0$, $T_{\pm} = T$, $k_\pm = k/4$, $\mu^+_R=\mu_R$, $\mu^-_L=\mu_L$ and set $c_+ = c_- = \frac{c}{2}$. Then, we find that
	\begin{equation}
		J_+^x+J_-^x = \frac{k}{4\pi} \left(\mu_L-\mu_R\right)
		\qquad
		J_+^x - J_-^x = -\frac{k}{4\pi} \left(\mu_L+\mu_R \right)
		\qquad
		T^{tx} = \frac{k}{8\pi} \left(\mu_L^2 - \mu_R^2\right)
	\end{equation}
again in perfect agreement with \cite{bd2}.
\item{{\bf Velocity driven steady state}}
	If we take $\mu^+_R=\mu^-_L=0$, $T_\pm=T$ and set $c_+ = c_- = \frac{c}{2}$ we find
	\begin{equation}
		T^{tx} = \frac{c \pi }{12} T^2 \frac{\beta_R}{1+\beta_R}
	\end{equation}
\end{itemize}

A large class of $d$-dimensional CFTs has an exact dual (holographic) description in terms of classical gravity in $d+1$ dimensions. Given the universal nature of our results, one may ask if it is possible to construct the gravity duals of these steady state flows. In order to holographically confirm our formula for the steady state currents in terms of the parameters of the asymptotic heat baths, we should ideally construct a supergravity solution for the full time dependent non-equilibrium configuration with the two wavefronts (steps) pushing outward into the heat baths. Solving the full time dependent Einstein equations typically requires numerics. An analytic approach is provided by the fluid-gravity correspondence \cite{Bhattacharyya:2008jc}. For any given stress-tensor of the boundary field theory, the fluid-gravity correspondence gives a systematic framework of how to construct the dual bulk geometry order by order in a derivative expansion. In general dimension this approach of course is only applicable if the system is in the hydrodynamic regime.

For $1+1$ dimensional CFTs, in the absence of a conserved charge, the situation simplifies dramatically. Einstein's equations for the dual three dimensional gravity fix the spacetime to be (locally) empty AdS$_3$. Even for a time-dependent, far from equilibrium flow the full gravity solution should be related by a large coordinate transformation to static AdS. Indeed it was shown in \cite{Haack:2008cp} that for the three dimensional spacetime dual of a $1+1$ dimensional CFT the expansion of the fluid-gravity correspondence truncates after the leading order. In fact, the full gravity solution dual to an arbitrary conformal stress tensor of the form in eq.~\ref{cartesianstress} has been given in \cite{Haack:2008cp}. By construction, this gravity dual realizes the universal uncharged steady state currents found here.

So far we have considered configurations in which the system under consideration is infinite in extent. Before proceeding let us consider a system of finite size. For simplicity we will restrict ourselves to uncharged configurations, though all the statements we will make in this context can be easily transferred to configurations with conserved charges.

\begin{figure}[hbt]
\begin{center}
\includegraphics[width=150mm]{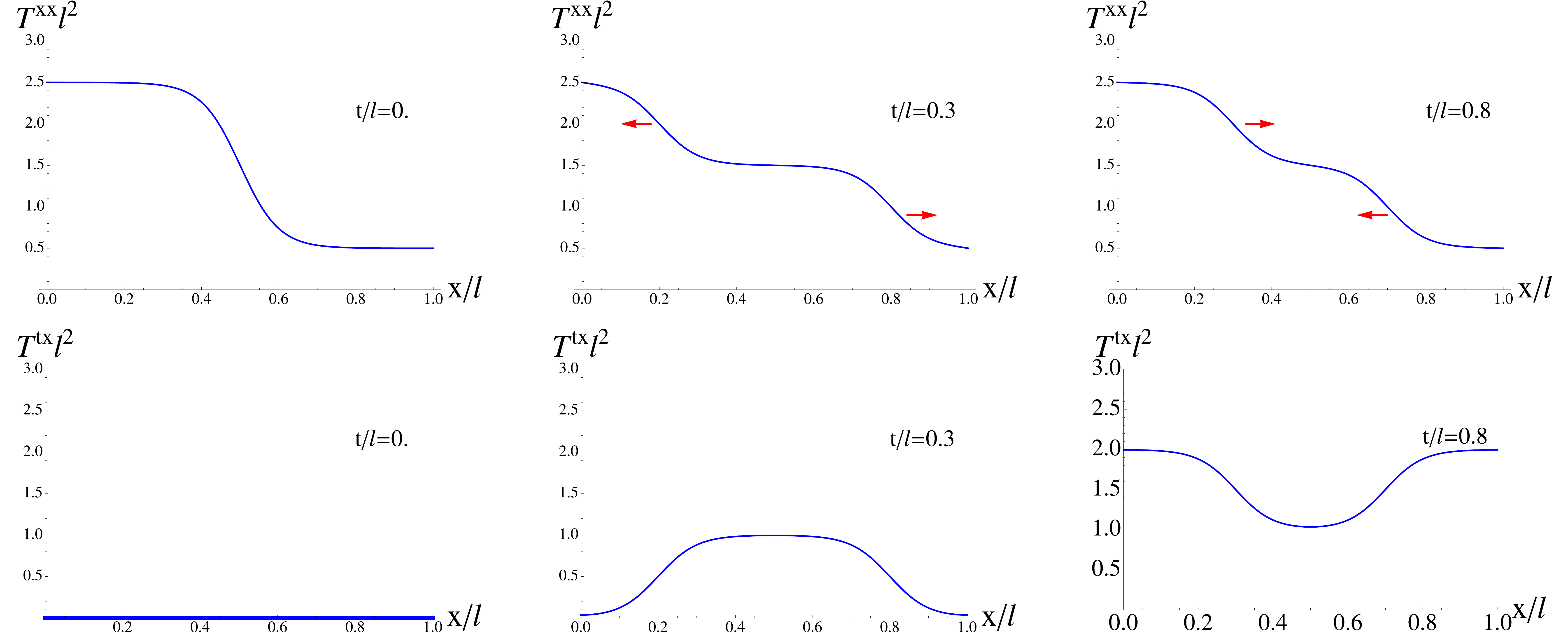}
\caption{\label{F:2dcftbox} The appearance of an intermediate steady state in a two dimensional finite sized system which is destroyed once the wavefronts (or steps) reach the edge of the system. The pressure is depicted in the top row. Around $t=0.5 \ell$ the wavefronts (whose direction is depicted by the red arrows) reach the edge of the system after which the steady state is destroyed. While the pressure and energy density exhibit periodic behavior the heat current, depicted in the lower row increases without bound. One can observe the emergence of a meta-steady state at intermediate times.}
\end{center}
\end{figure}

Following equation \eqref{cartesianstress} we observe that the $T^{tt}=T^{xx}$ components of the stress tensor satisfy a wave equation. The unique solution to a wave equation is fixed by an initial profile, its time derivative and boundary data for all time. Here, this data is provided by the value of $T^{tt}=T^{xx}$ at our initial time $t=0$, $T^{tx}$ at $t=0$ and the boundary value of $T^{tt}$ at, say, $x=0$ and $x=\ell$, the spatial extent of the box. Note that, as opposed to the case of an infinite system, the value of the heat flux at the boundaries $T^{tx}(x=0)$ and $T^{tx}(x=\ell)$ is fixed by the dynamics.
More practically, we specify an initial temperature and heat flow profile, or equivalently two functions $T_+(x)$ and $T_-(x)$, defined on an interval $0 \leq x \leq \ell,$ which are compatible with the boundary data $T_+(t)+T_-(-t) = P_L$ and $T_+(t+\ell)+T_-(-t+\ell)=P_R$.
In terms of this data we find that $T_+$ and $T_-$ are given for all $t$ and $x$ by
\begin{align}
\begin{split}
	T_+(x) &= \begin{cases} -n (P_L - P_R) + T_+(x_0) & x_0>0 \\
					     -(n-1)(P_L-P_R) + (P_R-T_-(-x_0) ) & x_0<0
		      \end{cases} \\
	T_-(x) &= \begin{cases} -n (P_L - P_R) + T_-(x_0) & x_0>0 \\
					     -(n-1)(P_L-P_R) + (P_R-T_+(-x_0)) & x_0<0
		      \end{cases}
\end{split}
\end{align}
where $x = x_0 + 2n\ell$, $-\ell<x_0<\ell$ and $n$ is an integer.
Strictly speaking, this finite sized system exhibits no steady-state (e.g., $T_{tx}(t,x=0)$ will increase without bound). In this particular case one might be tempted to relate the absence of the steady state to the lack of viscosity in two dimensional conformally invariant systems. We will see shortly that viscosity is not the only ingredient required to generate a steady state in a finite sized configuration. However, regardless of whether a true steady state forms at late times, if the system is large enough relative to the size of the initial configuration and we look at times smaller than the size of the system (in units of the speed of light) then we will see an intermediate steady state emerging. This is the manifestation of the steady state configuration described above in a finite sized system. A picture is worth a thousand words---see figure \ref{F:2dcftbox}.

\section{Universal flows in more than two dimensions}
\label{S:higherd}

We are interested in generalizing the 1+1 dimensional analysis of the previous section to a higher space-time dimension.
For simplicity, from now on we will restrict ourselves to the case of neutral fluids and therefore consider only heat transport. Though, much of what we say should be easy to generalize to charged fluids.
Obviously, the structure of having a superposition of a purely left moving wave and a purely right moving wave as the most general solution to the conservation equations for the stress tensor crucially relied on the special features of conformal symmetry in $1+1$ dimensions. We  argue that in higher dimensions this same type of physical behavior still correctly captures the late time dynamics of a generic initial energy-momentum profile connecting two infinitely far heat baths.

As a warm-up we study a system whose dynamics are described by linearized hydrodynamics. In this particular case we show explicitly that the dynamical behavior of the stress-tensor is, once again, described by a left-moving wave and a right-moving wave pushing into the equilibrated asymptotic regions with a steady state plateau generated in between. That is, at late times the steady state solution in the intermediate regime is connected to the asymptotic heat baths by two step-like features moving at the speed of sound and extending the steady state region. We then argue that, more generally, postulating a configuration which takes this two-step form, energy and momentum conservation alone (without any reference to the actual dynamics of the system as encapsulated by the constitutive relations of hydrodynamics) fixes the properties of the emerging steady state in terms of the pressures of the left and right heat baths. We then contend that under mild assumptions the ansatz may apply to non relativistic configurations as well. We conclude this section with numerical evidence which supports our ``steps ansatz''.

\subsection{Linearized hydrodynamics.}
\label{SS:LHydro}

Consider the stress tensor of ideal hydrodynamics
\begin{equation}
	T^{\mu\nu} = \epsilon(P) u^{\mu} u^{\nu} + (\eta^{\mu\nu} +u^{\mu}u^{\nu} ) P  + \mathcal{O}(\partial)\,.
\end{equation}
The equations of motion are given by
\begin{equation}
	\partial_{\mu}T^{\mu\nu} = 0\,.
\end{equation}
Obviously $P=P_0$ and $u^{\mu} = (1,0)$ with constant $P_0$ is a solution. Let us consider the following linearized perturbation: 
\begin{equation}
	P = P_0 + \delta P(t,x)
	\qquad
	\beta = \delta \beta(t,x)
\end{equation}
where $u^{\mu} = \gamma(1,\,\beta,\,0,\ldots,\,0)$ and $\gamma^{-2} = 1-\beta^2$.

In this limit of small perturbations, energy-momentum conservation amounts to a coupled wave equation for the pressure and velocity field whose solution is given by
\begin{align}
\begin{split}
	\delta P &= P_-(x-c_s t)+P_+(x+c_s t)\\
	\delta \beta(t,x) &= \beta_0 + \frac{1}{c_s(P_0 + \epsilon(P_0))} \left(P_+(x+ c_s t)- P_-(x-c_s t)\right)\,,
\end{split}
\end{align}
with $c_s^{-2} = \frac{\partial \epsilon}{\partial P}$ the speed of sound squared.
Placing heat baths at spatial infinity amounts to imposing
\begin{equation}
	T^{\mu\nu} = \hbox{diagonal}\left( \epsilon(P_0 \pm \Delta P) ,\, (P_0 \pm \Delta P) ,\, \ldots  ,\, (P_0 \pm \Delta P) \right)
\end{equation}
at $x\to \mp \infty$,
i.e., we have a heat bath with pressure $P_0+\Delta P$ on the left and a heat bath with pressure $P_0-\Delta P$ on the right. As long as $\Delta P$ is small our linearized hydrodynamic solution should be valid. Applying these boundary conditions we get that at $t\to\infty$
\begin{equation}
	T^{\mu\nu} =
		\begin{pmatrix}
			\epsilon (P_0) & \frac{ \Delta P}{c_s} & 0 & \ldots &  \\
			\frac{ \Delta P}{c_s} & P_0 & 0 & \ldots &  \\
			0 & 0 & P_0 & \ldots & \\
			0 & 0 & 0 & \ldots &  \\
			\vdots & \vdots & \vdots & \ldots &  \\
			0 & 0 & 0 & \ldots & P_0 \\
		\end{pmatrix}\,.
\end{equation}
In particular, we can read off the steady state heat current
\begin{equation}
	T^{t x}\Big|_{t\to\infty} =  \frac{\Delta P}{c_s}.
\end{equation}
Note that, just as in 1+1 dimensions, this result is independent on the details of the spatial temperature and velocity profile at $t=0$.

\subsection{An ansatz for steady state configurations}
\label{S:ansatz}
We have seen that in 1+1 dimensions generic initial conditions connected to asymptotic heat baths lead to a very simple picture for the development of a stationary state. The same picture also emerges in any dimension within the regime of validity of linearized inviscid hydrodynamics. In the central region a stationary plateau develops, connected by left and right moving waves to the asymptotic heat baths on either side. In order to ensure conservation of energy and momentum, the waves push outwards at a constant speed, expanding the size of the steady state plateau. In 1+1 dimensions the waves moved at the speed of light, whereas in linear hydrodynamics they moved at the speed of sound (which formally equals the speed of light in 1+1 dimensions).

These observations lead us to conjecture that, for generic CFTs connected to heat baths at infinity, the exact same structure will emerge irrespective of the dynamical details. At late times the energy-momentum tensor will be characterized by a right moving and left moving wave which are not necessarily hydrodynamic. These push into the asymptotic regions at fixed speeds $v_L$ and $v_R$, not necessarily equal to the speed of sound. In the central region between the waves an intermediate steady state regime will emerge. With this form of late time behavior, conservation of energy and momentum alone determine the heat current in the stationary state in terms of the asymptotic temperatures $T_L$ and $T_R$. More precisely, the resulting non-linear equations allow for two distinct branches of solutions each of which is completely determined by the relative pressure differences of the heat bath. Up to this $\mathbf{Z}_2$ ambiguity, the expression for the heat current is universal in that it only relies on the properties of the asymptotic regions and is  insensitive to any dynamical details of the system. For example, if we study a fluid that is described by viscous hydrodynamics including the full non-linear structure, the steps ansatz gives us a steady state heat current that is insensitive to the value of the shear viscosity. In fact, most of our analysis goes through even if we relax the condition of our field theory being conformal. We'll work with a general equation of state for the equilibrium system until the late stages of our analysis, where we will finally focus on the conformal case in order to get explicit expressions for the heat current.

Let us assume that at late times we can distinguish between three spatial regions as follows:
\begin{itemize}
\item{Region I} $-\infty < x < -L$ (left moving wave)
\item{Region II} $-L < x < L$ (steady state)
\item{Region III} $ L < x< \infty$ (right moving wave)
\end{itemize}
where we take $L$ to be very large and to grow with time so that at $t\rightarrow\infty$  all we have is region II. In all regions the equations of motion are energy-momentum conservation $\partial_{m}T^{mn} = 0$. The boundary conditions we use are
\begin{equation}
\label{E:Rbath}
	T^{\mu\nu}(x\to-\infty) = \begin{pmatrix}
		\epsilon( P_0 + \Delta P ) & 0 \\ 0 & P_0 + \Delta P
		\end{pmatrix}
\end{equation}
and
\begin{equation}
\label{E:Lbath}
	T^{\mu\nu}(x\to+\infty) = \begin{pmatrix}
		\epsilon (P_0 - \Delta P) & 0 \\ 0 & P_0 - \Delta P
		\end{pmatrix}
\end{equation}
where $\mu,\nu=t,x$ and the remaining coordinates of the stress tensor are on the diagonal.

Our ansatz is such that in the left region (region I) the solution takes the form
\begin{equation}
	T^{\mu\nu}_I = \begin{pmatrix}
		-\frac{1}{v_L} W_L (x + v_L t) + \epsilon (P_0+\Delta P) & W_L(x+v_L t) \\
		W_L(x+v_L t) & -v_L W_L(x+v_L t) + (P_0+\Delta P)
	\end{pmatrix}
\end{equation}
where \eqref{E:Lbath} implies that
\begin{equation}
	W_L(-\infty)=0
	\qquad
	W_L(\infty) = J_L\,.
\end{equation}
In the right region we have
\begin{equation}
	T^{\mu\nu}_{III} = \begin{pmatrix}
		\frac{1}{v_R} W_R (x - v_R t) +\epsilon (P_0 - \Delta P) & W_R(x - v_R t) \\
		W_R(x-v_R t) & v_R W_R(x-v_R t) + (P_0-\Delta P)
	\end{pmatrix}
\end{equation}
with
\begin{equation}
	W_R(\infty)=0
	\qquad
	W_R(-\infty) = J_R\,.
\end{equation}
In the central, steady-state, region we find from energy-momentum conservation that $T^{tx}_{II}$ and $T^{xx}_{II}$ are constant. Thus,
\begin{equation}
	T_{II}^{\mu\nu} = \begin{pmatrix}
		T_{II}^{tt}(x) & J \\
		J & P_{II}
		\end{pmatrix}\,.
\end{equation}

Now let us match the solutions such that at late times the right side of the left moving wave and the left side of the right moving wave determine the asymptotics of the central region. That is,
\begin{equation}
		T^{\mu\nu}_I (t\to\infty) =
		T^{\mu\nu}_{II}(x\to-\infty)
	\qquad
		T^{\mu\nu}_{III} (t\to\infty) =
		T^{\mu\nu}_{II}(x\to\infty)
\end{equation}
The resulting equations are			
	\begin{align}
	\begin{split}
		\begin{pmatrix}
			-\frac{J_L}{v_L} +\epsilon(P_0+\Delta P) & J_L \\
			J_L & -v_L J_L + (P_0+\Delta P)
		\end{pmatrix}
		&=\begin{pmatrix}
		T_{II}^{tt}(-\infty) & J \\
		J & P_{II}
		\end{pmatrix}
	\\
		\begin{pmatrix}
			\frac{J_R}{v_R} +\epsilon(P_0-\Delta P) & J_R \\
			J_R & v_R J_R + (P_0-\Delta P)
		\end{pmatrix}
		&=\begin{pmatrix}
		T_{II}^{tt}(\infty) & J \\
		J & P_{II}
		\end{pmatrix}
\end{split}
\end{align}
whose solution is:
\begin{align}
\begin{split}
\label{E:Matching}
	J &= J_L = J_R=\frac{2\Delta P}{v_L+v_R} \\
	P_{II} &= P_0 - \frac{v_L-v_R}{v_L+v_R} \Delta P \\
	T^{tt}(\infty) &= \frac{J}{v_R} + \epsilon (P_0-\Delta P) \\
	T^{tt}(-\infty) &= -\frac{J}{v_L} + \epsilon (P_0+\Delta P) \,.
\end{split}
\end{align}
Within our ansatz the steady state solution is parameterized by two constants $v_L$ and $v_R$ and an unknown function $T^{tt}(x)$ whose boundary data is known. To proceed we need some information on the constitutive relations which allow us to write $T^{tt}$ in terms of $T^{tx}$ and $T^{xx}$.

Let us suppose that the system is in thermal equilibrium where the steps connect to the steady state region. That is, at the interface between region $I$ and $II$ and at the interface between region $II$ and $III$ we have
\begin{equation}
\label{E:idealc}
	T^{\mu\nu} = \epsilon(P) u^{\mu}u^{\nu} + P (u^{\mu} u^{\nu} + \eta^{\mu\nu})\,.
\end{equation}
Imposing that $T^{\mu\nu}_{II}(\pm\infty)$ satisfies \eqref{E:idealc}  provides us with three equations at $+\infty$ and three equations at $-\infty$ with six unknown parameters: the two pressures and velocities at the ends of the left and right steps respectively constitute four parameters, and the left and right velocities $v_L$ and $v_R$. Note that up to this point we have made no reference to whether our system is conformal or not. However, without an explicit expression for $\epsilon(P)$ the equations are implicit and difficult to solve. So let us  specialize to the conformal case where $\epsilon(P) = (d-1)P$. In the present context this relation amounts to the statement that for a thermally equilibrated fluid moving at constant velocity,
\begin{equation}
\label{E:idealc2}
	T^{tx} = \pm\frac{\sqrt{((d-1)T^{tt}-T^{xx})((d-1)T^{xx}-T^{tt})}}{d-2}\,.
\end{equation}
Note that the reality condition on $T^{tx}$ implies that not all stress tensors will allow for an ideal hydrodynamic constitutive relation.

The solution for $v_L$ and $v_R$ can be written in a relatively simple parametric form. First, it is useful to switch to the following dimensionless variables:
\begin{equation}
	\delta p= \frac{\Delta P}{P_0}
	\qquad
	v_L = (x+\delta p) \nu
	\qquad
	v_R = (x-\delta p) \nu
\end{equation}
In principle, $\delta p$ is an external parameter that we tune and one would want to solve for $v_L$ and $v_R$ (or $\nu$ and $x$) in terms of $\delta p$. Such an expression can be obtained explicitly: one finds a simple equation for $\nu^2$ as a function of $x$ and $\delta p$ but the equation for $x$ is a sixth order polynomial which can be written as  a product of a quadratic polynomial and a fourth order one and therefore solvable. The solution can be written in closed form but is not very illuminating.Instead, we find it more useful to write the solution in parametric form.

\begin{figure}[hbt]
\begin{center}
\includegraphics{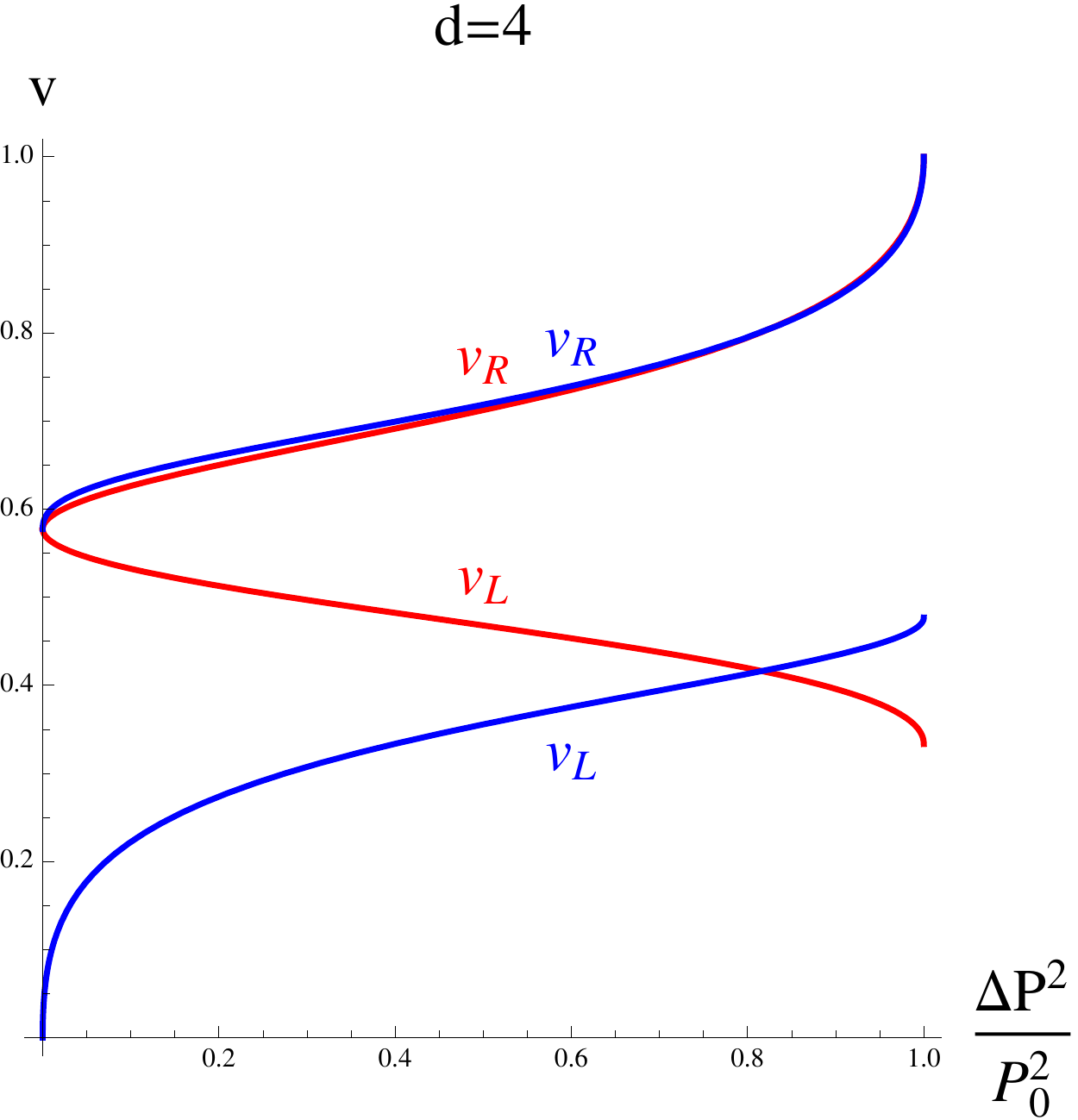}
\caption{\label{F:vplots} The velocity of the left and right moving steps as a function of the relative pressure difference. There are two branches of solutions. The red branch which we refer to as the thermodynamic branch, coincides in the zero pressure difference limit with ideal hydrodynamics. The blue branch is somewhat of a mystery.}
\end{center}
\end{figure}

We find that there are two branches of solutions. The first branch is given by
\begin{align}
\begin{split}
\label{E:thermobranch}
	(\delta p)^2 &=  \frac{d x(4-d(2+x))}{(d-2)^2} \\
	\nu^2 &= -\frac{(d-2)^2}{2(d-1)x(d(d-2)+(2+d(d-2))x)} \,,
\end{split}
\end{align}
where $-1+\frac{2}{d} < x<0$. In this branch $v_L$ decreases monotonically from $1/{\sqrt{d-1}}$ to $1/(d-1)$ and $v_R$ increases from $1/\sqrt{d-1}$ to $1$. We will refer to this branch as the ``thermodynamic branch''.\footnote{%
After this work appeared on arxiv the authors of \cite{Bhaseen:2013ypa} have managed, using a particular change of variables, to simplify the form of the parametric solution of the thermodynamic branch to $v_L = \frac{1}{d-1} \sqrt{\frac{\sqrt{\frac{1+\delta p}{1-\delta p}}+(d-1)}{\sqrt{\frac{1+\delta p}{1-\delta p}}+(d-1)^{-1}}}$ and $v_R^{-1} =  \sqrt{\frac{\sqrt{\frac{1+\delta p}{1-\delta p}}+(d-1)}{\sqrt{\frac{1+\delta p}{1-\delta p}}+(d-1)^{-1}}} $. 
}
The second branch is given by
\begin{align}
\begin{split}
	( \delta p)^2 & = \frac{(d-2)^2 x^2 (d^2 - 4 d (d-1)x - (d-2)^2 x^2 )}{P(x)} \\
	\nu^2 & = \frac{P(x)}{4 (d-1) x^2 (2 x+ (d-2) d (1+x))^2}
\end{split}
\end{align}
with
\begin{equation}
	P(x) = d (d(d-2)^2 + 4d(d-1)(d-2)x+(-16+d(32+d(-24+7 d)))x^2)\,.
\end{equation}
where $\tilde{x}_+ < x < 0$ with $\tilde{x}_+ = -\frac{d(d-1)-\sqrt{d(8+d(-11+4 d))}}{(d-2)^2}$. Here $v_R$ starts at $1/\sqrt{d-1}$ and ends at 1 but $v_L$ starts at zero and ends at $v_L^2 = \frac{2+(d-2)d(2d-1)}{2(d-1)^4} - \frac{d-2}{2(d-1)^4}\sqrt{d(8-11d+4 d^2)}$. A plot of these two solutions in four space-time dimensions can be found in figure \ref{F:vplots}.

Within our two-step ansatz we have found two possible universal expressions for the steady state heat current connecting two asymptotic heat baths. In 1+1 dimensions and in d+1 dimensional linear hydrodynamics we were able to show that this 2-step solution is indeed the late-time configuration for any system irrespective of details of the initial data. The interesting question that remains at this stage is whether our 2-step solution has a similar status as a universal late time configuration for a generic d+1 dimensional CFT. In section \ref{SS:numerical} we will demonstrate that this is the case at least within systems well approximated by viscous hydrodynamics by numerically constructing solutions to non-linear hydrodynamical equations of motion. Before doing so we briefly discuss an extension of our step ansatz to non relativistic systems.

\subsection{Generalizations}

We may break our analysis in the previous subsection into two parts. The ``steps'' ansatz gave us \eqref{E:Matching} almost immediately. We then needed to commit to an equation of state and assume that the matching can be done in a region where an ideal fluid equation of state holds to get our final result. The validity of \eqref{E:Matching} can also be argued for on more intuitive grounds. 
Let us write the momentum conservation equation in the form
\begin{equation}
\label{E:conservation}
\partial_t \pi + \partial_x P =0
\end{equation}
where $\pi$ is the momentum density in the $x$-direction and $P$ is the pressure. Equation \eqref{E:conservation} is valid for relativistic as well as non relativistic systems. If we assume that after a long time $t$ we reach a steady state with constant $\pi=\pi_{late}$ in a plateau region of size $(v_L+v_R) t$ and 
non-vanishing in a region whose sized doesn't grow with $t$,
then integrating the momentum conservation equation over $x$ and $t$ gives us
\begin{equation}
\pi_{late} = - \frac{\left . P \right |^{x \rightarrow \infty}_{x \rightarrow - \infty}}{v_L + v_R}.
\end{equation}
To reach \eqref{E:Matching} we need to use the symmetry properties of the relativistic stress tensor, $T^{tx} = T^{xt}$, which imply that heat flow and momentum density are the same. While this relation is in general violated in non-relativistic systems, one can make a similar statement in incompressible non-relativistic fluids. For such fluids the particle number current $j$ is related to the momentum density $\pi$ via
\begin{equation}
j = \pi/\rho,
\end{equation}
where $\rho$ is the mass density.
If we further impose non-relativistic conformal invariance,
which constrains stress-energy tensor via $dT^{xx} = 2 T^{tt}$ where $d$ is the number of spatial dimensions (see e.g., \cite{LandL5,Son:2005rv,Adams:2008wt})
our steps ansatz gives a universal relation for the particle number current in terms of the asymptotic properties of the heat bath,
\begin{equation}
(\Delta \rho j)_{late} =  - \frac{\left . 2 T^{tt} \right |^{x \rightarrow \infty}_{x \rightarrow - \infty}}{v_L + v_R}.
\end{equation}
In order to obtain $v_L$ and $v_R$ need to commit to a particular equation of state.
\subsection{Numerical evidence for the ansatz}
\label{SS:numerical}

To support our ansatz we provide numerical evidence for its reliability within the realm of hydrodynamics. Hydrodynamics may be thought of as an effective theory, valid for systems which are close to thermodynamic equilibrium. The proximity to thermodynamic equilibrium is  quantified by the smallness of the derivative of thermodynamic parameters. More precisely, in the absence of a conserved charge the thermodynamic parameters are temperature $T$ and the center of mass velocity $u^{\mu}$ (normalized so that $u^{\mu}u_{\mu}=-1$). Relativistic hydrodynamics provides equations of motion for $T$ and $u^{\mu}$ when these are slowly varying in space and time. These equations of motion are energy conservation $\partial_{\mu}T^{\mu\nu}=0$ supplemented with a set of constitutive relations: a local expression for the stress tensor in terms of the thermodynamic parameters and their gradients. For a conformal theory, to second order in the gradient expansion, we have \cite{Baier:2007ix,Bhattacharyya:2008jc}
\begin{equation}
\label{E:Tdef}
	T^{\mu\nu} = P (d u^{\mu} u^{\nu} + \eta^{\mu\nu}) +\pi^{\mu\nu}
\end{equation}
with
\begin{equation}
\label{E:pidef}
	\pi^{\mu\nu} = - \eta \sigma^{\mu\nu} + \eta \tau_{\pi} \Sigma_0^{\mu\nu} + \lambda_i \Sigma_i^{\mu\nu}
\end{equation}
and
\begin{align}
\begin{split}
	\sigma^{\mu\nu} &= 2 \partial^{\langle \mu}u^{\nu\rangle} \qquad
	\Sigma_0^{\mu\nu} =u^{\langle \alpha}\partial_{\alpha} \sigma^{\mu\nu\rangle} + \frac{1}{d-1} \sigma^{\mu\nu} \partial_{\alpha}u^{\alpha} \\
	\Sigma_1^{\mu\nu} & = \sigma^{\langle\mu}{}_{\lambda}\sigma^{\lambda\nu\rangle} \qquad
	\Sigma_2^{\mu\nu}  = \sigma^{\langle\mu}{}_{\lambda}\omega^{\lambda\nu\rangle} \qquad
	\Sigma_3^{\mu\nu}  = \omega^{\langle\mu}{}_{\lambda}\omega^{\lambda\nu\rangle}
\end{split}
\end{align}
where we use triangular brackets to denote a transverse traceless symmetric contraction
\begin{equation}
	A_{\langle \mu \nu \rangle} = \frac{1}{2} P_{\mu}{}^{\alpha}P_{\nu}{}^{\beta} \left(A_{\alpha\beta}+A_{\beta\alpha}\right) - \frac{1}{d-1} P_{\mu\nu}P^{\alpha\beta} A_{\alpha\beta}
\end{equation}
and have used $\omega_{\mu\nu} = \frac{1}{2}P^{\alpha}{}_{\mu}P^{\beta}{}_{\nu} \left(\partial_{\alpha}u_{\beta} - \partial_{\beta}u_{\alpha}\right)$ for the vorticity tensor. The dependence of the pressure, shear viscosity, and other transport coefficients on the temperature are completely fixed by the conformal symmetry. e.g.,
\begin{align}
	P &= p_0 T^d &
	\eta &= d p_0 T^{d-1} \eta_0 \\
	\tau_{\pi} & = \tau_0 T^{-1} &
	\lambda_1 &= \zeta_1 T^{d-2}\,.
\end{align}
Similar expressions hold for $\lambda_2$ and $\lambda_3$. Note that $\eta_0 = \eta/s$.

The equations of motion $\partial_{\mu}T^{\mu\nu}=0$ are non-linear equations for the variables $u^{\mu}$ and $T$. Without the second order terms, $\Sigma^{\mu\nu}$, these equations are non causal, unstable and, in certain instances, ill-defined \cite{PhysRevD.31.725}. By including the second order terms in the constitutive relations one obtains a well defined initial value problem \cite{HL}. The numerical scheme which we will adopt for solving the equations of motion can be found in \cite{Luzum:2008cw} (See also \cite{Teaney:2009qa} for a detailed review). After reinterpreting \eqref{E:pidef} as an equation of motion for $\pi_{\mu\nu}$,
\begin{align}
\begin{split}
\label{E:pikinematic}
	\pi^{\mu\nu} =& -\eta\sigma^{\mu\nu} - \tau_{\pi} \left( u^{\langle \alpha}\partial_{\alpha} \pi^{\mu\nu\rangle} + \frac{d}{d-1}\pi^{\mu\nu} \partial_{\alpha}u^{\alpha}\right)  \\
	&+\frac{\lambda_1}{\eta^2} \pi^{\langle \mu}{}_{\lambda} \pi^{\lambda\nu\rangle} - \frac{\lambda_2}{\eta}
\pi^{\langle\mu}{}_{\lambda}\omega^{\lambda\nu\rangle} + \lambda_3 \omega^{\langle\mu}{}_{\lambda}\omega^{\lambda\nu\rangle}
\end{split}
\end{align}
we obtain a coupled set of equations for the components of $\pi^{\mu\nu}$ (from \eqref{E:pikinematic}) the velocity field $u^{\mu}$ and the temperature $T$ (from $\partial_{\mu}T^{\mu\nu}=0$ together with \eqref{E:Tdef}).

In order to verify the steady state ansatz described in section \ref{S:ansatz} we look for solutions to the hydrodynamic equations of motion where the hydrodynamic fields depend only on time and one spatial coordinate ($x$) and are fixed such that the system is in thermal equilibrium at spatial infinity. In this setup the velocity field has one component which we parametrize via $u^{\mu} = (\cosh(\kappa(t,x)),\,\sinh(\kappa(t,x)),\,0\,\ldots,\,0)$. Note that this implies that, in our setup, the vorticity tensor vanishes, $\omega^{\mu\nu}=0$. The dissipative tensor $\pi^{\mu\nu}$ has only one relevant independent component which we choose to be $\pi_{xx}(t,x)$,
\begin{equation}
	\pi_{tt} = \pi_{xx} \tanh^2\kappa\,,
	\qquad
	\pi_{tx} = \pi_{xx} \tanh \kappa\,.
\end{equation}

Let us denote $V = (T,\,\kappa,\,\pi_{xx})$, then \eqref{E:pikinematic} together with $\partial_{\mu}T^{\mu\nu}=0$ and \eqref{E:Tdef} take the schematic form
\begin{equation}
\label{E:EOM}
	\partial_t V = M V
\end{equation}
where $M$ denotes a matrix valued differential operator involving only the spatial coordinate $x$. Since equation \eqref{E:EOM} takes a canonical form it can be solved using standard numerical techniques. We've used pseudo-spectral methods to compute the spatial derivatives $M V$  (see, for instance, \cite{trefethen2000spectral,boyd2001chebyshev}). For time evolution we've used either (classical) fourth order Runge-Kutta or Adams Bashforth. In order to implement boundary conditions at spatial infinity we've used a finite interval to parameterize the spatial coordinate (i.e., $y = L \tanh(x/L)$).

\begin{figure}[hbt]
\begin{center}
\includegraphics[width=140mm]{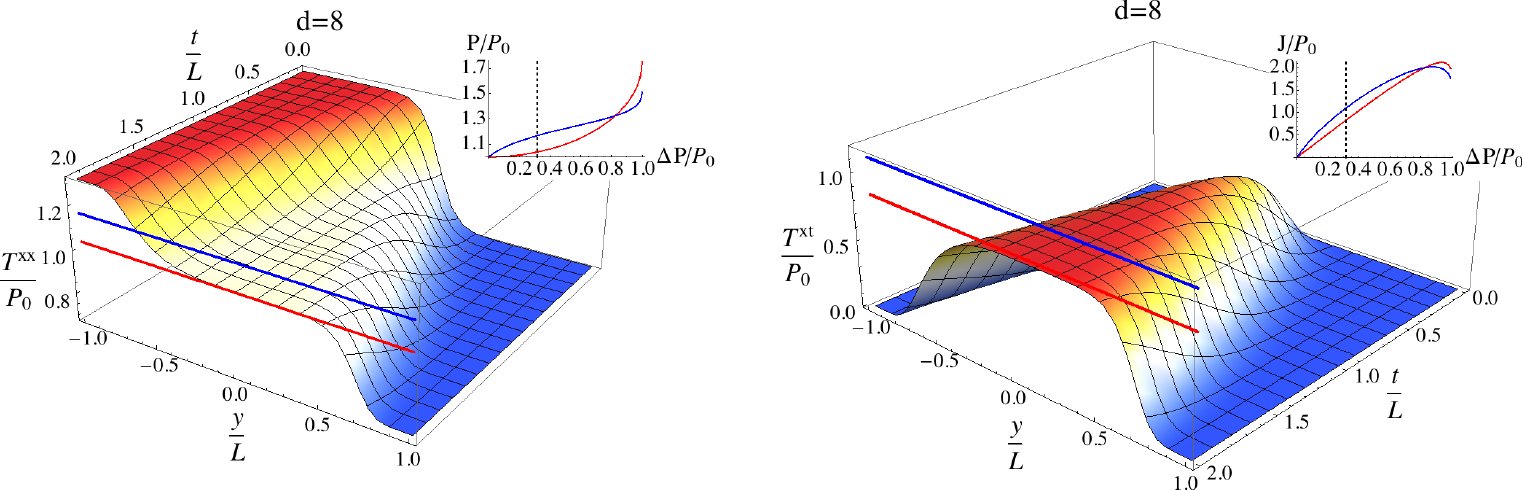}
\includegraphics[width=140mm]{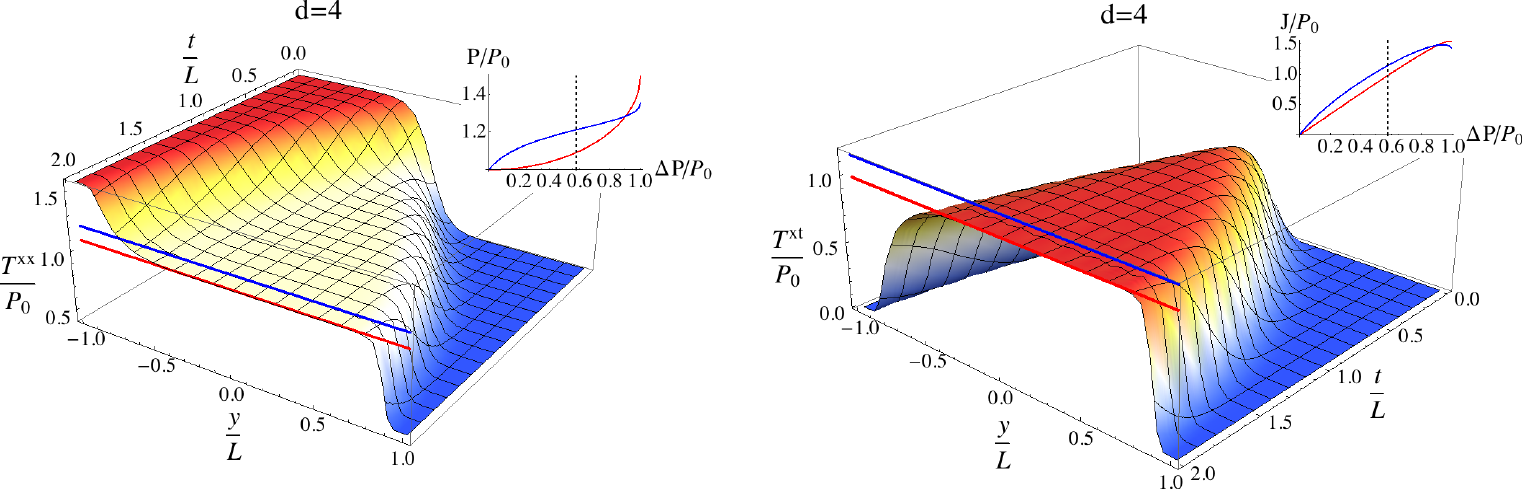}
\includegraphics[width=140mm]{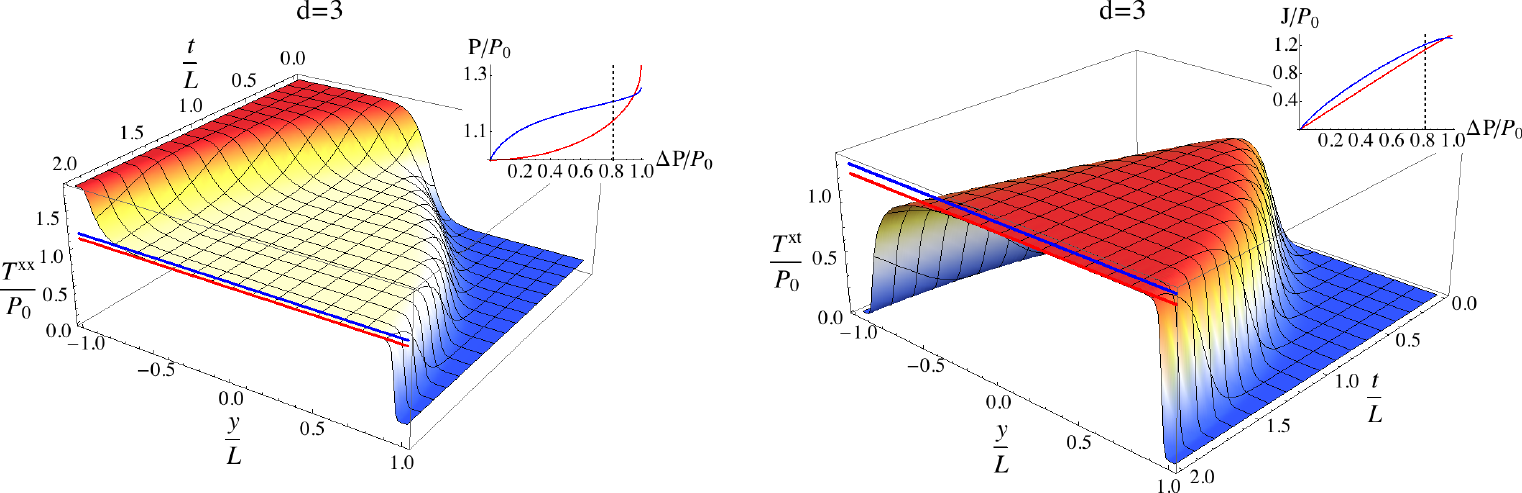}
\caption{\label{F:quenches} Plots of the $tx$ and $xx$ components of the energy momentum tensor as a function of time for $d=3,\,4$ and $8$ space-time dimensions for various values of the pressure difference of the heat baths. In all three instances the initial pressure profile was controlled by a $\tanh$ function in the compactified $y$ coordinates. The pressure difference between the heat baths is denoted $\Delta P$ and the average pressure is $P_0$. The prediction of the steps ansatz is signified by the red and blue lines also given in the inset. The values used for the transport coefficients in these plots are characterized by $\eta_0 = 1/40\pi$, $\tau_0 = 1/20\pi$ and $\lambda_0=0$.}
\end{center}
\end{figure}

\begin{figure}[hbt]
\begin{center}
\includegraphics[width=140mm]{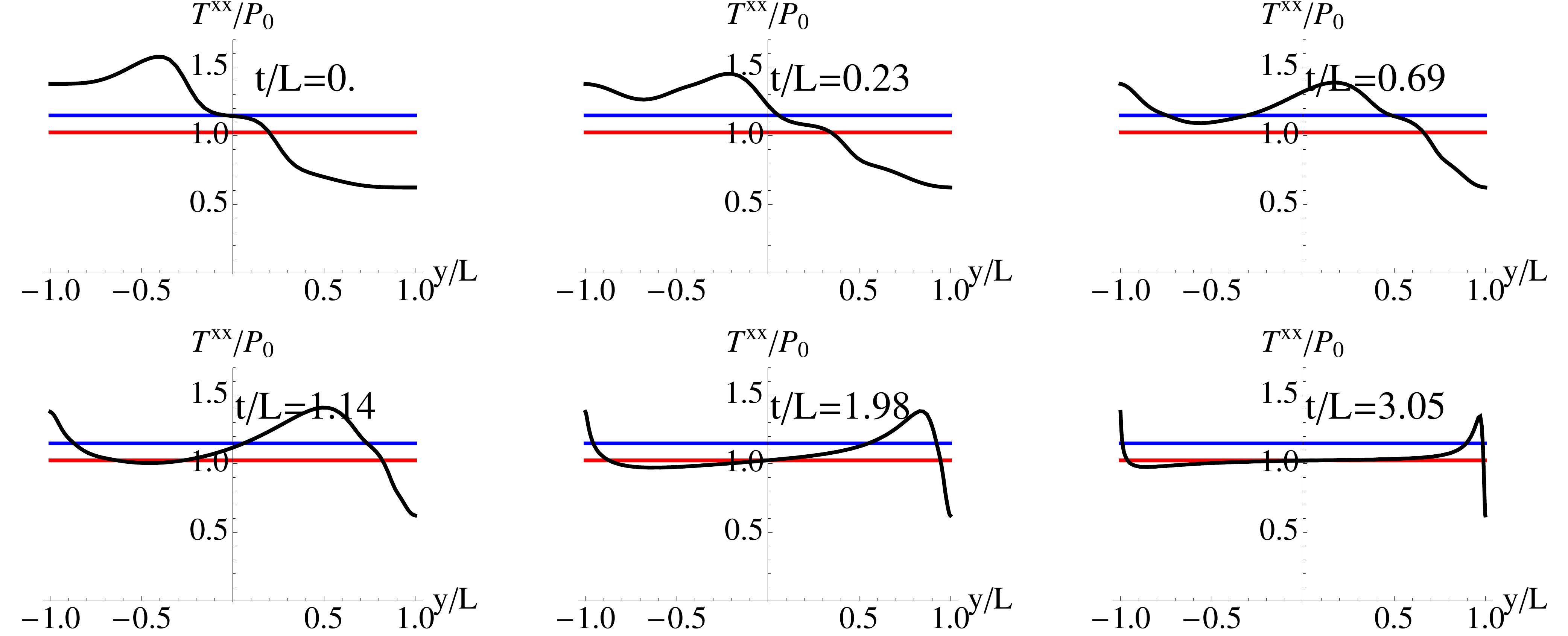}
\caption{\label{F:other} Plot's of the $T^{xx}$ component of the stress tensor at various times for a non quench-like initial condition. For lack of space we have not depicted the corresponding heat fluxes which are non zero at the initial time. At around $t=2L$ the initial disturbance has been pushed to infinity and a steady state emerges. The prediction of the steps ansatz is signified by the red and blue lines. The values for the transport coefficients in these plots is given by $\eta_0 = 1/8\pi$ $\tau_0=(2-\ln 2)/4\pi$ and $\lambda_0=0$.}
\end{center}
\end{figure}

In our preliminary analysis we have managed to check the validity of the steps ansatz up to, roughly $\Delta P^2 / P_0^2 = 2/3$ for initial conditions which resemble a spatial quench. See figure \ref{F:quenches}.
At large $\Delta P^2 / {P}_0^2$ the numerical result deviated from our steady state solution \eqref{E:thermobranch} to within roughly 0.1\% in  the time scales we worked with for which we had reasonably stable solutions. In figure \ref{F:other} we have plotted the time evolution for initial configurations which are not quench-like and obtain similar results.

\section{Discussion}
\label{sec:conclude}

In this work we have argued for the existence of a steady state for configurations which involved infinitely separated heat baths. If we place our system in a finite sized box then one needs to be somewhat more careful about the final steady-state configuration. For a two dimensional conformal fluid we have seen that the steady-state configuration which we are proposing is valid at times which are large compared to the initial length scale of our profile divided by the speed of light, but at time scales small compared to the size of the box divided by the speed of light.

Indeed, even if the ``steps ansatz'' is valid at intermediate times, once the waves of the steps ansatz crash into the sides of the box the steps ansatz ceases to be valid and the system will start to settle down to its ultimate late time solution. Here one has to distinguish between two cases. In the absence of dissipation, e.g. in two dimensions or in linear inviscid hydrodynamics, the pressure and heat current profile will slosh back and forth in the box forever and no steady state will exist.
With dissipation one would expect the system to increase its entropy by dissipative processes and to eventually reach a diffusion dominated steady state. For instance, in relativistic viscous hydrodynamics there is a unique steady state flow, as worked out for example in \cite{Khlebnikov:2010yt}, connecting a heat bath of temperature $T_L$ to a heat bath of temperature $T_R$. 
The properties of the end steady-state depend in detail on the value of the shear viscosity.

Thus, we expect the steps ansatz to be valid in any configuration whose dynamics are not fully dominated by diffusion---at late enough times non diffusive processes will generate a flow driven steady state whose behavior is universal. In other words,  generically, one would expect a system with an initial profile which interpolates between two heat baths to realize two separate steady states: at intermediate times a flow driven universal steady state of the type we describe here and at late times the standard diffusion driven steady state which depends on the particular details of the theory.

It should be emphasized that neither of these steady states is guaranteed to always be available. The late time solution to linear viscous hydro in a box is that the pressure is linear in the spatial position and constant in time but the velocity field is linear in time and constant in space. This observation is only relevant in some ``pretend world" where linear viscous hydro is always valid. In practice once the amplitude of the perturbations becomes large the system will be driven into a non linear regime and there, presumably a diffusion driven steady-state does exist. The ``flow driven" steady state need not always exist either.  As we mentioned previously, if the dynamics of a system is well modeled by a diffusion equation alone, the steps ansatz is not a solution.

For a relativistic conformal theory we have worked out explicitly the universal expression for the expectation value of the heat current for our flow dominated steady-state. To be precise, the ansatz we use predicts two possible branches of solutions which we refer to as the thermodynamic branch and the other branch. Therefore there are two possible steady-state heat currents. While our current numerics indicate that the thermodynamic branch is the preferred steady-state solution (See in particular figure \ref{F:other}), our search was less than exhaustive. It may well be that the final steady-state currents will depend on the dynamics of the system under study, or, that the system prefers the branch where the heat current is smallest. In the latter case the other branch will be chosen at large relative pressure difference. We hope to report on such matters in the near future.

Agreement of Israel-Stewart hydrodynamics with the steady-state ansatz is relatively good but leaves room for improvement especially for large relative pressure difference $\delta p = \Delta P / P_0$. Obtaining solutions for very late times implies using grids much larger than the ones we can currently work with in a reasonable time frame. This is partly due to technical difficulties which arise when compactifying the parametric length of the spatial coordinate. However, it could be that our hydrodynamic description breaks down once the relative pressure difference is of order unity. When $\delta p$ is of order 1, gradients in the system become large and the dissipative tensor $\pi_{\mu\nu}$ introduced in section \ref{SS:numerical} becomes large breaking the validity of the derivative expansion. Worse, when the dissipative tensor $\pi_{\mu\nu}$ becomes large Israel Stewart theory breaks down due to negative entropy production \cite{Baier:2007ix,Teaney:2009qa}. Thus, one should view our numerical construction as preliminary evidence for the validity of our steps ansatz. More satisfying evidence for the validity of our ansatz could follow from dynamics which are better adopted to both large and small pressure differences. The gauge-gravity duality might be the perfect arena for carrying out such computations \cite{Chesler:2013lia}. Better yet, since our ansatz is applicable to $3+1$ dimensional systems, one might hope to obtain experimental evidence for its reliability.

\section*{Acknowledgments}
We'd like to thank the organizers of the Gauge/Gravity Duality 2013 workshop at the MPP in Munich where this work was initiated. Special thanks to Joe Bhaseen for introducing us to the topic. We would also like to thank D. Podolsky for useful discussions, M. Rozali and W. van der Schee for helpful tips on numerics and K. Jensen for initial collaboration on this project. The work of HC and AK is supported, in part, by the US Department of Energy under grant number DE-FG02-96ER40956. AY is a Landau fellow, supported in part by the Taub foundation. AY is also supported by the ISF under grant number 495/11, by the BSF under grant number 2014350, by the European commission FP7, under IRG 908049 and by the GIF under grant number $1156/2011$.

\bibliographystyle{JHEP}
\bibliography{Mixed}

\end{document}